\RequirePackage{lineno}

\documentclass[aps,prc,showpacs,showkeys,preprintnumbers,amsmath,amssymb,superscriptaddress]{revtex4}

\usepackage{lineno}

\usepackage{graphicx} 
\usepackage{amssymb}
\usepackage{dcolumn}
\usepackage{bm}
\usepackage{color}
\newcommand{ \be }{\begin{linenomath*}\begin{eqnarray}}
\newcommand{ \ee }{\end{eqnarray}\end{linenomath*}}
\newcommand{ \la }{\langle}

\newcommand{ \ra }{\rangle}

\def \mean#1 {{\la #1 \ra}}

\newcommand*{\Scale}[2][4]{\scalebox{#1}{$#2$}}%
\definecolor{dgreen}{cmyk}{1.,0.,1.,0.2}        
\definecolor{orange}{cmyk}{0.,0.353,1.,0.}    

\begin{document}
\title{The Role of  Baryon Number Conservation in Measurements of Fluctuations}
\author{Claude A. Pruneau}
\email{aa7526@wayne.edu}
\affiliation{Department of Physics and Astronomy, Wayne State University, Detroit, MI 48201, USA}
\date{\today}

\begin{abstract}
I discuss the role and impact of net-baryon number conservation in measurements of net proton fluctuations in heavy-ion collisions. I show that the magnitude of the fluctuations is entirely determined by the strength of two particle correlations. At LHC and top RHIC energy, this implies the fluctuations  are proportional to  the integral of the balance function (BF), $B^{p\bar p}$ of protons and anti-protons, while in the context of the RHIC beam energy scan (BES), one must also account for correlations of ``stopped" protons. The integral of $B^{p\bar p}$ measured in a $4\pi$ detector depends on the relative cross-sections of processes
yielding $p\bar p$ and  those  balancing the proton baryon number via the 
production of other anti-baryons. The accepted integral of $B^{p\bar p}$ further depends on
the shape and width of the BF relative to the width of the acceptance. The magnitude
of the measured second order cumulant of  net proton fluctuations thus has much less to do  with QCD susceptibilities than with the creation/transport of baryons and anti-baryons in heavy-ion collisions, and most particularly the impact of radial flow on the width of the BF.
I thus advocate that  net-proton fluctuations should be studied by means of differential BF measurements rather  than with integral correlators. I also 
derive an expression of net-baryon fluctuations in terms of integrals of balance functions of identified baryon pairs and argue that measurements of such balance functions would enable a better understanding of the collision system expansion dynamics, the hadronization chemistry, and an experimental assessment of the strength of net-baryon fluctuations. 
\end{abstract}
\maketitle

\section{Introduction}

Lattice QCD calculations (LQCD)  with physical quark masses suggest that at RHIC top energy and LHC energy, the matter produced in heavy-ion collisions consists of a state of matter known as Quark Gluon Plasma (QGP)~\cite{Adams:2005dq,RHICqgp2}.  LQCD also indicates that for vanishing baryon chemical potential ($\mu_B$), the transition from the QGP to a hadron gas  phase (HGP) is of crossover type~\cite{Aoki:2006we}, while at large baryon chemical potential, it should be of first order. This implies the existence of a critical point (CP).  Theoretical considerations further suggest that within the vicinity of the CP, one should expect sizable changes in the matter's correlation length and that divergent net-charge ($\Delta Q$), net-strangeness ($\Delta S$), or net-baryon ($\Delta B$) fluctuations should occur~\cite{PhysRevLett.102.032301}. Away from the CP, in the cross-over region, some trace of critical behavior might also 
remain~\cite{Aoki:2006we,Karsch:2000ps}. There is thus a strong  interest in mapping the magnitude of $\Delta Q$, 
$\Delta S$, $\Delta B$ with $\mu_B$ and temperature ($T$). This can be accomplished, in principle, by measuring
second, third, and fourth order cumulants of these quantities as a function of beam energy ($\sqrt{s_{\rm NN}}$). 
However, a number of caveats must be considered.   First, LQCD predicts the magnitude of $\Delta B$ fluctuations in a finite coordinate space volume, $V$,   but, experimentally, in heavy-ion collisions, these  are measured  based on a specific volume, $\Omega$,  in momentum space. It is ab initio unclear how charge transport (e.g., flow, diffusion, etc), within the QGP produced in heavy-ion collisions map $V$ onto $\Omega$ and how this mapping shall affect the fluctuations observed in momentum space~\cite{KITAZAWA201565}. Second, $\Delta B$ fluctuations  in $V$ are not globally constrained by  net-baryon number conservation while those in $\Omega$ intrinsically are~\cite{BRAUNMUNZINGER2016805}. Third, it is not obvious that  a measurement of proton vs. anti-proton fluctuations is sufficient to make a statement about baryon number fluctuations. What is indeed the effect of the unobserved baryons, i.e., anti-neutron ($\bar n$), anti-lambda ($\bar \Lambda$), etc? A host of other questions may also be considered, including whether the produced system has time to thermalize globally and whether, consequently, it is meaningful to invoke the notion of susceptibility.

In this paper, I first focus the discussion on  fluctuations of conserved charges, more specifically the net-proton number $\Delta N_p$, and examine the impact of baryon number conservation on measurements of the
second cumulant $\kappa_2(\Delta N_p)$. I next consider the effects  of a partial 
measurement of baryon fluctuations based on fluctuations of the net proton number.   Finally, I  extend the discussion 
and consider fluctuations of net-baryons in terms of contributions from identified pairs of baryons and anti-baryons.

In the context of the Grand Canonical Ensemble (GCE), 
fluctuations of $\Delta B$ are related to the 
reduced susceptibility $\hat \chi^{B}_2$  according to~\cite{PhysRevC.96.024910,BRAUNMUNZINGER2016805}
\be\label{eq:susceptibility}
\hat \chi^B_2  = \frac{1}{VT^3} \kappa_2(\Delta B),
\ee
where $V$ is the volume of the system, $T$ its temperature, and $\kappa_2(\Delta B)$, the second 
 order cumulant of  $\Delta B$. The second order cumulants  amounts to  the variance and is calculated according to 
\be
\kappa_2(\Delta B) = \la \Delta B^2\ra - \la \Delta B\ra^2,
\ee
where $\la \Delta B\ra$ and $\la \Delta B^2\ra$ are the first and second moments, measured over an ensemble of events, of the net-baryon number   $\Delta B = N_B - N_{\bar B}$. The variables   $N_B$ and $N_{\bar B}$ represent  multiplicities of baryon and anti-baryons, respectively,  within the volume $V$   in a particular instance of the system (collision). Averages are computed over all possible instances of the system. Within the GCE, the susceptibility  $\hat \chi^B_2$ is calculated as the second derivative of the reduced thermodynamic pressure $\hat p = p/T^4$ w.r.t. the reduced baryon chemical potential $\hat \mu_B \equiv \mu_B/T$
\be
\hat \chi^B_2 = \frac{\partial^2 \hat p}{\partial \hat \mu_B^2}.
\ee
Higher cumulants, $\kappa_n(\Delta B)$, $n\ge 3$,  of the net-baryon number $\Delta B$ are likewise related to 
higher order susceptibilities corresponding to $n$-th derivatives   of the pressure. Because these susceptibilities have a finite dependence on the volume of the system, which is relatively ill defined in the context of nucleus-nucleus collisions, it is customary to consider ratios of the cumulants $\kappa_n(\Delta B)$ by $\kappa_2$ to eliminate this dependence. 
Higher cumulants are deemed of great interest  because of their higher power dependence on the correlation length $\xi$ which should diverge in the vicinity of the CP~\cite{Athanasiou:2010kw}. 

Measurements of  second, third, and fourth order cumulants of
$\Delta Q$, $\Delta S$, and  $\Delta B$ have been conducted at RHIC, in particular, in the context of the first beam energy scan (BES I)~\cite{Kumar:2013cqa,Sharma:2015jdk}. Cross-cumulants have also been reported~\cite{Chatterjee:2016mve}. While second, third, and fourth order of 
$\Delta Q$ and $\Delta S$   are observed to have either modest or monotonic  dependence on the beam energy, the  third and fourth 
cumulant of the net proton number exhibit   non-monotonic behaviors vs. $\sqrt{s_{\rm NN}}$, with 
what appears to be a statistically significant minimum near $\sqrt{s_{\rm NN}} =20$ GeV. 
Interestingly, this energy is also the locus of a minimum in  the magnitude of directed flow, $v_1$,
observed in Au--Au collisions vs. beam energy~\cite{Singha:2016mna}. The existence of these two minima at the same
energy has been interpreted as an indicator of the presence of the CP in this vicinity~\cite{Mohanty_2011}. However, the observed non-monotonic behavior and minimum
have  received a variety of other interpretations~\cite{PhysRevC.98.054901}. Indeed, several caveats
may impact the interpretation of the existing   results, as well as those
of   future experiments. Primary
among these are concerns  associated with the role of baryon number conservation. 

The total baryon number of an isolated system is a conserved quantity. This implies  that the net-baryon number of all particles produced 
in a given A--A collision   should add to the sum
of the baryon numbers of the incoming nuclei. However,  fluctuations of the net-baryon number, $\Delta B$, shall  be observed
when measuring baryon production in a fiducial 
 acceptance limited to central rapidities. This much is true.    Furthermore, it is  generally assumed that the measured
 magnitude of $\kappa_n(\Delta B)$ shall inform us about about the susceptibilities $\chi_n^B$. It is 
  argued, in particular, that great care has to be given to the choice of the width of the rapidity acceptance used in measurements
 of $\kappa_n(\Delta B)$: too narrow an acceptance should  lead to Poisson fluctuations of $\Delta B$ while 
 too wide an acceptance should greatly suppress the fluctuations because the net 
 baryon number of the entire system must be conserved. Moreover, it is often  stated  
that for an acceptance of about one to two  units of rapidity, such as those of the STAR and ALICE experiments, 
the effect of  baryon number conservation should  be negligible  and only small corrections need to be applied
to interpret  $\kappa_n(\Delta B)$ measurements in terms of susceptibilities.   Unfortunately, these assertions 
are factually incorrect as I shall demonstrate in this paper: at LHC and top RHIC energies, the non-trivial part of the cumulant $\kappa_2(\Delta B)$ is entirely determined 
by baryon number conservation and the width of the experimental acceptance, while at lower energies of 
the RHIC Beam-Energy-Scan (BES), one must account for fluctuations in the proton yield associated with baryon stopping and collision geometry. The good news, however, is that  local baryon number conservation applies both in infinite static matter and   within a system (heavy-ion collision) 
undergoing  fast longitudinal and radial expansion.  The only important consideration then is how radial and longitudinal 
expansion affect the fraction of conserved baryons focused within the experimental acceptance, on average.
While such a fraction cannot be measured directly by means of cumulants, it can be assessed and 
extrapolated, in principle, from measurements of balance functions. It is my goal, in  this paper,
to  demonstrate that second cumulants of the net-baryon number are intrinsically and entirely determined 
by baryon number conservation, radial flow,  and the width of the acceptance. I further show
that  while integral correlators, such as $\kappa_2(\Delta B)$,  are sensitive to radial flow, they do not allow
easy discrimination between effects of radial flow and the width of the acceptance in transverse momentum, $p_{\rm T}$, and pseudorapidity, $\eta$. 
However,  differential correlation functions in the form of balance function (BF)   offer a 
much better method to assess the interplay between finite acceptance, radial flow, and baryon number conservation. 

In order to demonstrate these assertions, I first need to express the second order cumulant of net-baryon (proton)  fluctuations, measured within a specific acceptance, in terms of second order (pair) factorial cumulants. I will then show that these are 
related to the $\nu_{\rm dyn}$ correlation observable, which in turn, is proportional to the integral, within the same acceptance, of the baryon balance function. I will show how the integral  of the balance function
is determined by the hadro-chemistry of the collision system and that the shape and width of the balance function are largely determined by longitudinal and radial flow.

This paper is divided as follows. Section~\ref{sec:Definitions}  defines moment, cumulant, factorial moment,  factorial cumulant and balance function notations  used in the remainder of the paper. The Poisson limit 
of fluctuations and the relation between $\kappa_2$ and the  $\nu_{\rm dyn}$ correlator are discussed in Sec.~\ref{sec:SkellamLimit}. The connection between
$\nu_{\rm dyn}$ and the balance function, and the role of baryon number conservation at LHC and top RHIC energy 
are discussed in Sec.~\ref{sec:LHCLimit}, while the impact of baryon stopping and a net excess of baryons in 
the fiducial volume of the measurement are addressed in Sec.~\ref{sec:Stopping}. Section~\ref{sec:NetBaryonFluctuations} extends the discussion 
of net-baryon fluctuations in terms of a sum of  balance functions of identified baryon and anti-baryon pairs. 
 Conclusions are  summarized in Sec.~\ref{sec:Summary}.

%
\section{Definitions and Notations}
\label{sec:Definitions}
\subsection{Integral Correlators} 
\label{sec:Integral Correlators}

For  simplicity,  all particles of interest (e.g., protons and anti-protons) are  assumed to be measured in the same fiducial momentum acceptance $\Omega$. Measured  multiplicities of species $\alpha$ and $\beta$,  in a given event, are denoted $N_{\alpha}$ and $N_{\beta}$, respectively. 
Anti-baryons are indicated with over-bar symbols, e.g., $N_{\bar\alpha}$ denotes the multiplicity of anti-particles of species $\alpha$.
For instance, proton and anti-proton multiplicities are  denoted $N_{p}$ and $N_{\bar p}$, where as the 
net-proton number is written  $\Delta N_p=N_p-N_{\bar p}$. 

Theoretically, the fluctuations may be described in terms of a joint probability $P(N_{\alpha}, N_{\beta}|\Omega, C)$ determined by the acceptance $\Omega$ and the  centrality $C$ of the heavy-ion collisions 
of interest. Experimentally, fluctuations may be characterized in terms of moments of multiplicities calculated
as  event ensemble averages denoted $\la O\ra$.   First and second moments of multiplicities $N_{\alpha}$ and $N_{\beta}$  are defined according to
\be
m_1^{\alpha} &=& \la N_{\alpha} \ra = \sum_{i=0}^{\infty} N_{\alpha} P(N_{\alpha}, N_{\beta}|\Omega, C), \\
m_2^{\alpha,\beta}&=& \la N_{\alpha}N_{\beta} \ra = \sum_{i=0}^{\infty} N_{\alpha}N_{\beta} P(N_{\alpha}, N_{\beta}|\Omega, C). 
\ee
Cumulants of multiplicities $N_{\alpha}$ and $N_{\beta}$ are written
\be
\label{eq:kappa1}
\kappa_1^{\alpha} &=&  m_1^{\alpha}, \\ 
\label{eq:kappa2}
\kappa_2^{\alpha,\beta} &=&  m_2^{\alpha,\beta}- m_1^{\alpha}m_1^{\beta}. 
\ee
The cumulants $\kappa_2^{\alpha,\alpha}$ and $\kappa_2^{\alpha,\beta}$, with $\beta \ne \alpha$,  correspond to the variance of $N_{\alpha}$ and the covariance of $N_{\alpha}$ and $N_{\beta}$, respectively. 

Experimentally, particle losses associated with the detection and event reconstruction modify these moments and cumulants. 
Corrections for such losses are most straightforward when carried out for single particles and pairs of particles. It is thus
convenient to introduce factorial moments of the multiplicities $N_{\alpha}$ and $N_{\beta}$ as
\be
f_1^{\alpha} &=& \la N_{\alpha} \ra = m_1^{\alpha}, \\ 
f_2^{\alpha,\beta} &=& \la N_{\alpha}  N_{\beta} -\delta_{\alpha,\beta} N_{\alpha}\ra =m_2^{\alpha,\beta} - \delta_{\alpha,\beta} m_1^{\alpha}.
\ee
Given  factorial moments of measured multiplicities $n_{\alpha}$ and $n_{\beta}$, corrected factorial moments are obtained as 
\be
f_1^{\alpha} &=& \tilde f_1^{\alpha}/\varepsilon_{\alpha}, \\ 
f_2^{\alpha,\beta} &=& \tilde  f_2^{\alpha,\beta}/(\varepsilon_{\alpha}\varepsilon_{\beta}),
\ee
where $\tilde f_1^{\alpha}$ and $\tilde f_2^{\alpha,\beta}$ represent raw (or uncorrected) factorial moments, while $\varepsilon_{\alpha}$ and $\varepsilon_{\beta}$ are detection efficiencies for particle species $\alpha$ and $\beta$, respectively. Note 
that best experimental precision may require one accounts for
dependences of these quantities on the transverse momentum, the azimuth angle, and the pseudorapidity of the particles~\cite{PhysRevC.96.054902,PhysRevC.98.014905}.
By construction, these factorial moments are determined by the single and pair densities of produced particles according to
\be
f_1^{\alpha} &=& \int_{\Omega} \rho_1^{\alpha}(\vec p) d^3p, \\ 
f_2^{\alpha,\beta}&=& \int_{\Omega} \rho_2^{\alpha,\beta}(\vec p_1,\vec p_2) d^3p_1d^3p_2, 
\ee
where $\rho_1^{\alpha}(\vec p)$ is the single particle density of particle species ${\alpha}$, 
and $\rho_2^{\alpha,\beta}(\vec p_1,\vec p_2)$
is the pair-density of particle species $\alpha$ and $\beta$.   

Factorial moments (corrected for efficiency losses)
are combined to obtain factorial cumulants according to 
\be
\label{eq:F1}
F_1^{\alpha} &=& f_1^{\alpha} = \kappa_1^{\alpha}  = m_1^{\alpha}, \\   \nonumber
F_2^{\alpha,\beta} &=& f_2^{\alpha,\beta} -  f_1^{\alpha}f_1^{\beta}, \\ 
\label{eq:F2}
& =& m_2^{\alpha,\beta}- \delta_{\alpha,\beta}m_1^{\alpha} -   m_1^{\alpha}m_1^{\beta}. 
\ee
Factorial cumulants   $F_2^{\alpha,\beta}$  are, by construction, true measures of pair correlations: they vanish identically in the absence of
particle correlations and take finite  values, either  negative or positive,  in the presence of such correlations. However,  null $F_2^{\alpha,\beta}$ values are not a 
sufficient condition to conclude measured particles are uncorrelated. Using 
the above definitions of first and second order factorial cumulants, one verifies second order cumulants may be written
\be\label{eq:kappa2vsF}
\kappa_2^{\alpha,\beta} &=& \delta_{\alpha,\beta}F_1^{\alpha} + F_2^{\alpha,\beta}. 
\ee

It is convenient to introduce normalized factorial cumulants defined according to 
\be\label{eq:R2}
R_2^{\alpha,\beta} &\equiv& \frac{ f_2^{\alpha,\beta} }{ f_1^{\alpha} f_1^{\beta}} - 1
= \frac{F_2^{\alpha,\beta} }{F_1^{\alpha} F_1^{\beta} },
\ee 
as well as the following linear combination of normalized two-cumulants:
\be\label{eq:nudyn}
\nu_{\rm dyn}^{\alpha,\beta} &=& R_2^{\alpha,\alpha} + R_2^{\beta,\beta} - 2 R_2^{\alpha,\beta},
\ee
where $\alpha \ne \beta$ represent two distinct types of particles.
The correlator $\nu_{\rm dyn}$ was originally introduced to search for the suppression of  net-charge fluctuations  in heavy-ion collisions~\cite{PhysRevLett.85.2076,Voloshin:1999yf,Pruneau:2002yf,PhysRevC.68.044905,PhysRevC.79.024906}.  It is of practical interest because it is experimentally robust, impervious to statistical fluctuations, and singles out dynamical fluctuations involved in particle production~\cite{Pruneau:2002yf}.  Its use has 
since been extended to study fluctuations of the relative yields of several  types of particle species 
at RHIC and LHC energies~\cite{Abelev:2009ai,Abelev:2012pv,Wang:2012jua,Acharya:2017cpf}.

\subsection{Balance Functions} 
\label{sec:Balance Functions}

General balance functions (BF) are differential correlations functions that contrast the strength of like-sign (in the context of this paper, same baryon number) and unlike-sign (opposite charge or opposite baryon number) particles correlations~\cite{PhysRevLett.85.2689,Bass:2001rc}. 
General balance functions for pairs of species $\alpha$ and $\beta$ are nominally defined according to
\be \label{eq:BF}
B^{\alpha,\bar \beta}(\Delta y) &=& 
\frac{1}{2} \left[
\frac{\rho_2^{\alpha,\bar \beta}(\Delta y)}{\rho_1^{\alpha}} 
- \frac{\rho_2^{\alpha,\beta}(\Delta y)}{\rho_1^{\alpha}} 
+  \frac{\rho_2^{\bar\alpha,\beta}(\Delta y)}{\rho_1^{\bar\alpha}} 
- \frac{\rho_2^{\bar\alpha,\bar\beta}(\Delta y)}{\rho_1^{\bar\alpha}}
\right],
\ee
in which labels without (e.g., $\alpha$, $\beta$) and with (e.g., $\bar\alpha$, $\bar\beta$) an over-bar  indicate baryons and anti-baryons, respectively. Expressions  $\rho_1^{\alpha}$  and and $\rho_2^{\alpha,\beta}(\Delta y)$   denote  single-particle and pair  densities of baryons (anti-baryons), respectively.  Particle $\alpha$ is considered as the ``trigger" or ``given" particle, while particle $\beta$ is regarded as  the ``associate". The ratios $\rho_2^{\alpha,\beta}(\Delta y)/\rho_1^{\alpha}$ are  conditional densities expressing the number of particles of species $\beta$ at a separation $\Delta y$ from a particle of species $\alpha$.  In the context of this work, it is useful to calculate the BF according to 
 \be\label{eq:BvsD2}
 B^{\alpha,\bar \beta}(\Delta y) = \frac{1}{2} \left[ D_2^{\alpha,\bar \beta }(\Delta y) + D_2^{\bar\alpha,\beta }(\Delta y) \right],
 \ee
 in which $D_2^{\alpha,\bar \beta }(\Delta y)$ and $D_2^{\bar\alpha,\beta }(\Delta y)$ represent differences of conditional densities defined as 
 \be
 D_2^{\alpha,\bar \beta }(\Delta y) &=& \rho_1^{\bar\beta} R_2^{\alpha,\bar\beta}(\Delta y)  - \rho_1^{\beta}  R_2^{\alpha,\beta}(\Delta y), \\ 
 D_2^{\bar\alpha,\beta }(\Delta y) &=& \rho_1^{\beta} R_2^{\bar\alpha,\beta}(\Delta y)  - \rho_1^{\bar\beta} R_2^{\bar\alpha,\bar\beta}(\Delta y),
 \ee
with normalized two-particle normalized cumulants
\be
\label{eq:R2Diff}
R_2^{\alpha,\beta}(\Delta y) = \frac{\rho_2^{\alpha,\beta}(\Delta y)}{\rho_1^{\alpha}\otimes \rho_1^{\beta}(\Delta y)} -1 = \frac{ F_2^{\alpha,\beta}(\Delta y)} { F_1^{\alpha} \otimes F_1^{\beta}(\Delta y)}.
\ee
The correlators $F_2^{\alpha,\beta}(\Delta y)=\rho_2^{\alpha,\beta}(\Delta y) -\rho_1^{\alpha}\otimes \rho_1^{\beta}(\Delta y)$ are differential factorial cumulants with an explicit dependence on the pair separation $\Delta y$. 


\section{Moments of Net Proton Distribution and Skellam Limit} 
\label{sec:SkellamLimit}

The net proton number is defined as $\Delta N_p \equiv N_p - N_{\bar p}$.
One straightforwardly verifies that its first and second cumulants  
are 
\be
\kappa_1(\Delta N_p) &=& \kappa_1^{p} - \kappa_1^{\bar p}, \\ 
\kappa_2(\Delta N_p) &=& \kappa_2^{p,p} + \kappa_2^{\bar p,\bar p} - 2 \kappa_2^{p,\bar p},
\ee
where the first and second cumulants of proton and anti-proton multiplicities, denoted
by the indices $p$ and $\bar p$, respectively, are defined according to Eqs.~(\ref{eq:kappa1},\ref{eq:kappa2}). 
These may alternatively  be written
\be
\label{eq:kappa1NetB}
 \kappa_1(\Delta N_p) &=& F_1^p - F_1^{\bar p}, \\ 
\label{eq:kappa2NetB}
 \kappa_2(\Delta N_p) &=& F_1^p + F_1^{\bar p} + F_2^{p,p} +  F_2^{\bar p,\bar p} - 2 F_2^{p,\bar p}.
\ee
%
One finds that the second cumulant of the net-proton number involves two parts, the first being determined 
by the average multiplicities of protons and anti-protons and a more interesting part driven by two-particle correlations.

As stated above, in the absence of two-particle or higher order particle correlations, the factorial moments $F_2^{\alpha,\beta}$ vanish. The 
Poisson limit of the second order cumulant, often called Skellam, is thus  simply
\be\label{eq:kapp2Skellam}
\kappa_2^{\rm Skellam}(\Delta N_p) &=& F_1^p + F_1^{\bar p}.
\ee
It is convenient to consider the ratio, $r_{\Delta N_p}$, of a measured cumulant $\kappa_2(\Delta N_p)$ and its Skellam limit. Using Eqs.~(\ref{eq:kappa2NetB},\ref{eq:kapp2Skellam}), one gets
\be\label{eq:kapp2Skellam2}
r_{\Delta N_p} \equiv \frac{\kappa_2(\Delta N_p)}{\kappa_2^{\rm Skellam}(\Delta N_p)}  = 1+\frac{F_2^{p,p} +  F_2^{\bar p,\bar p} - 2 F_2^{p,\bar p} }{F_1^p+F_1^{\bar p}}.
\ee
This may also be written
\be\label{eq:kapp2Skellam3}
r_{\Delta N_p}  = 1+
 \frac{  
\left(F_1^p\right)^2 R_2^{p,p} 
+ \left(F_1^{\bar p}\right)^2 R_2^{\bar p,\bar p}
-2 F_1^{p}F_1^{\bar p} R_2^{p,\bar p}
  }{F_1^p+F_1^{\bar p}},
\ee
where I inserted normalized factorial cumulants defined according to Eq.~(\ref{eq:R2}).

\section{LHC and Top RHIC Energy} 
\label{sec:LHCLimit}

At LHC and top RHIC energy, one has $\la N_p\ra \approx  \la N_{\bar p}\ra$. The ratio 
$r_{\Delta N_p}$ is thus approximately 
\be\label{eq:kapp2Skellam4}
r_{\Delta N_p}  &=& 1+\frac{F_1^p}{2} \left[  R_2^{p,p}  +  R_2^{\bar p,\bar p} -  2R_2^{p,\bar p}\right], \\ 
\label{eq:kapp2Skellam5}
&=& 1+\frac{1}{4}\la N_T\ra  \nu_{\rm dyn}^{p,\bar p},
\ee
where $\la N_T\ra = \la N_p\ra +\la N_{\bar p}\ra$ is formally  defined as 
\be
\la N_T\ra=  \int_{\Omega} \rho_1^{p}(\vec p) d^3p + \int_{\Omega} \rho_1^{\bar p}(\vec p) d^3p.
\ee 
But given the densities $\rho_1^{p}$ and $\rho_1^{\bar p}$ are approximately constant at central rapidities, one can write $\la N_T\ra=dN_T/d\eta \times \Delta \eta$, where $\Delta \eta$ represents the longitudinal width of the experimental acceptance. The ratio $r_{\Delta N_p}$ may thus be written 
\be\label{eq:SkellamRatio}
r_{\Delta N_p} = 1 + \frac{1}{4} \Delta \eta \frac{dN_T}{d\eta} \nu_{\rm dyn}^{p,\bar p}.
\ee
As I  discuss below, net-baryon number conservation implies that   $\nu_{\rm dyn}^{p,\bar p}$ is negative  with an absolute magnitude that  depends on the width  $\Delta \eta$ of the fiducial  acceptance. Neglecting this dependence, one  would expect the ratio $r_{\Delta N_p}$ to have a trivial, approximately linear, dependence on the width of the acceptance~\cite{BRAUNMUNZINGER2017114}:
\be\label{eq:rVsDeltaEta}
r_{\Delta N_p} \approx 1 - a \Delta \eta, 
\ee
where $a\equiv \frac{1}{4} dN_T/d\eta |\nu_{\rm dyn}^{p,\bar p}|$.  However, the value of  $\nu_{\rm dyn}^{p,\bar p}$
should itself depend on $\Delta \eta$. The above is thus likely to be a somewhat poor approximation of the actual dependence 
of $r_{\Delta N_p}$ on $\Delta \eta$.  I show later  in this section  that the quality of the approximation depends on the actual shape of the balance function and the rapidity range of interest. Additionally, given $r_{\Delta N_p}\rightarrow 1$ in the limit $\Delta \eta\rightarrow 0$, one might also be tempted to conclude that fluctuations of the net-proton number are Poissonian (Skellam) 
in that limit. That is actually incorrect. The true measure of correlations
is given by $\nu_{\rm dyn}^{p,\bar p}$, which is, in general,  non-vanishing  even in the limit $\Delta \eta\rightarrow 0$.  This is the case, e.g., for a system producing pions  via the decay of $\rho_0$-mesons (See e.g., Eq.~(81) of Ref.~\cite{Pruneau:2002yf}), and for systems that can be described with a balance function, as I demonstrate in the following.

It is clearly of interest to assess how the value of $\nu_{\rm dyn}^{p,\bar p}$ may 
depend on the acceptance of the measurement. This is readily achieved 
with the introduction of balance functions defined in Eqs.~(\ref{eq:BF},\ref{eq:BvsD2}). 
Using Eq.~(\ref{eq:BvsD2}), one finds that  proton-proton balance functions may be written 
\be
B^{p, p}(\Delta y) &=& \frac{1}{2} \left[ \rho_1^{\bar p} R_2^{p,\bar p}(\Delta y) - \rho_1^{p} R_2^{p,p}(\Delta y) +   \rho_1^{p} R_2^{\bar p,p}(\Delta y) - \rho_1^{\bar p} R_2^{\bar p,\bar p}(\Delta y)  \right], 
\ee
where $\rho_1^{p}$  and $\rho_1^{\bar p}$ are  single particle densities of protons  and  anti-protons, respectively, and 
$R_2^{p,p}(\Delta y)$, $R_2^{p,\bar p}(\Delta y)$, $R_2^{\bar p,p}(\Delta y)$, and  $R_2^{\bar p,\bar p}(\Delta y)$ are normalized cumulants of pair densities.  The variable $\Delta y = y_1 - y_2$ represents the difference between the rapidities  of particles $y_1$ and $y_2$ of any given pair.

In A--A collisions and in the limit  $\la N_p\ra =  \la N_{\bar p}\ra$, one has $ \rho_1^{\bar p}= \rho_1^{p}$
and  $R_2^{\bar p,p}(\Delta y) = R_2^{p,\bar p}(\Delta y)$. The BF  simplifies to
\be \nonumber
B^{p,p}(\Delta y) &=& - \frac{\Delta \eta}{4}  \frac{dN_T}{d\eta}  \left\{  R_2^{p,p}(\Delta y) \right. \\ 
& & + \left. R_2^{\bar p,\bar p}(\Delta y) - 2R_2^{p,\bar p}(\Delta y) \right\}.
\ee
Integration of  $F_2^{\alpha,\beta}(\Delta y)$ across the $\Delta y$ acceptance yields the integral factorial cumulant
$F_2^{\alpha,\beta}$ defined by Eq.~(\ref{eq:F2}). The integral of the BF can thus be written
\be
I_{p,\bar p}(\Omega) &=& - \frac{1}{4} \la N_T\ra  \nu_{\rm dyn}^{p,\bar p}(\Omega).
\ee
Up to a sign, the integral of the BF is equal to the second term of  Eq.~(\ref{eq:kapp2Skellam5}). One  can then write
\be\label{eq:OneMinusR}
1 - r_{\Delta N_p}  &=& I_{p,\bar p}(\Omega).
\ee
One concludes that at high-energy, i.e., in the limit $\la N_p\ra =  \la N_{\bar p}\ra$, the deviation of the Skellam ratio from unity  is identically equal to the integral of the BF.  It is thus useful to  examine what determines the magnitude of this integral.

Neglecting the effect of incoming and stopped protons from incoming projectiles (I account for these in Sec.~\ref{sec:Stopping}), the shape and amplitude of the BF  reflect how and where baryon-conserving balancing pairs  of protons and anti-protons are created and transported in the aftermath of A--A collisions.  If only an anti-proton ($Q=-1$,$B=-1$)  could balance the production of a proton ($Q=1$,$B=1$),   then, by construction,  the balance function would  integrate to unity
 over the full phase space of particle production. However, baryon number  
conservation can  be satisfied by the production of other anti-baryons.  
An  anti-baryon of some kind must indeed accompany the production of
a proton. The proton-baryon balance function may thus be written
\be
B^{p,\bar B}(\Delta y) = B^{p,\bar p}(\Delta y) + B^{p,\bar n}(\Delta y) + B^{p,\bar\Lambda}(\Delta y) + \cdots =  \sum_{\bar\beta} B^{p,\bar\beta}(\Delta y),
\ee
where the sum extends to all anti-baryons that can balance the production of a proton. By construction, 
this balance function must integrate to unity over the full particle production phase space:
\be\label{eq:IB4pi}
I_{p,\bar B}^{4\pi} &=& 1,
\ee
where $4\pi$ denotes that the integral is extending over all rapidities and transverse momenta.
The production of pairs $p\bar p$, $p\bar n$,  $p\bar \Lambda$, $p\bar \Sigma^-$, etc, have probabilities
determined  by their relative  cross-sections. These, in turn, must be equal to integrals of their
respective  balance functions. One can then write
\be
1 \equiv  I_{p,\bar B}^{4\pi} &=& I_{p,\bar p}^{4\pi} +  I_{p,\bar n}^{4\pi} + I_{p,\bar \Lambda}^{4\pi} + \cdots  = \sum_{\bar \beta} I_{p,\bar \beta}^{4\pi},
\ee
where, once again, $4\pi$ denotes that the integrals are extending over all rapidities and transverse momenta, and    $\sum_{\bar \beta}$ represents a sum over
all anti-baryons ($B=-1$). In this context, the functions $I_{p,\bar \beta}^{4\pi}$ can be considered 
as probabilities of the respective baryon number balancing processes determined by their cross-sections.
 The $p\bar p$ balance function integral  is  one of many components of the full $p,\bar B$ BF. Its value is thus smaller than unity. 
 \begin{figure}[htbp] 
   \centering
   \includegraphics[width=0.32\linewidth,keepaspectratio=true,clip=true,trim=10pt 0pt 5pt 5pt]{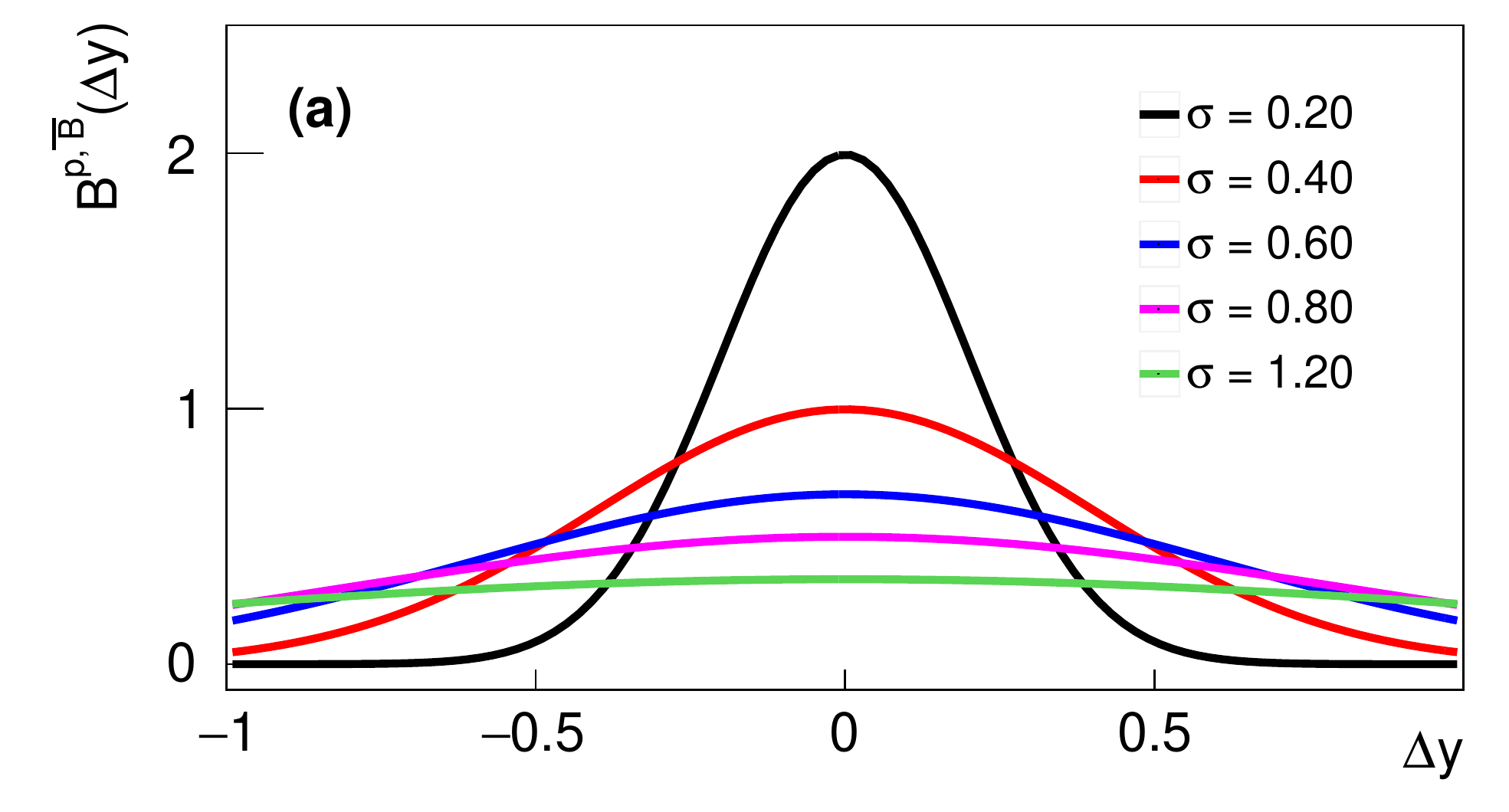} 
   \includegraphics[width=0.32\linewidth,keepaspectratio=true,clip=true,trim=10pt 0pt 5pt 5pt]{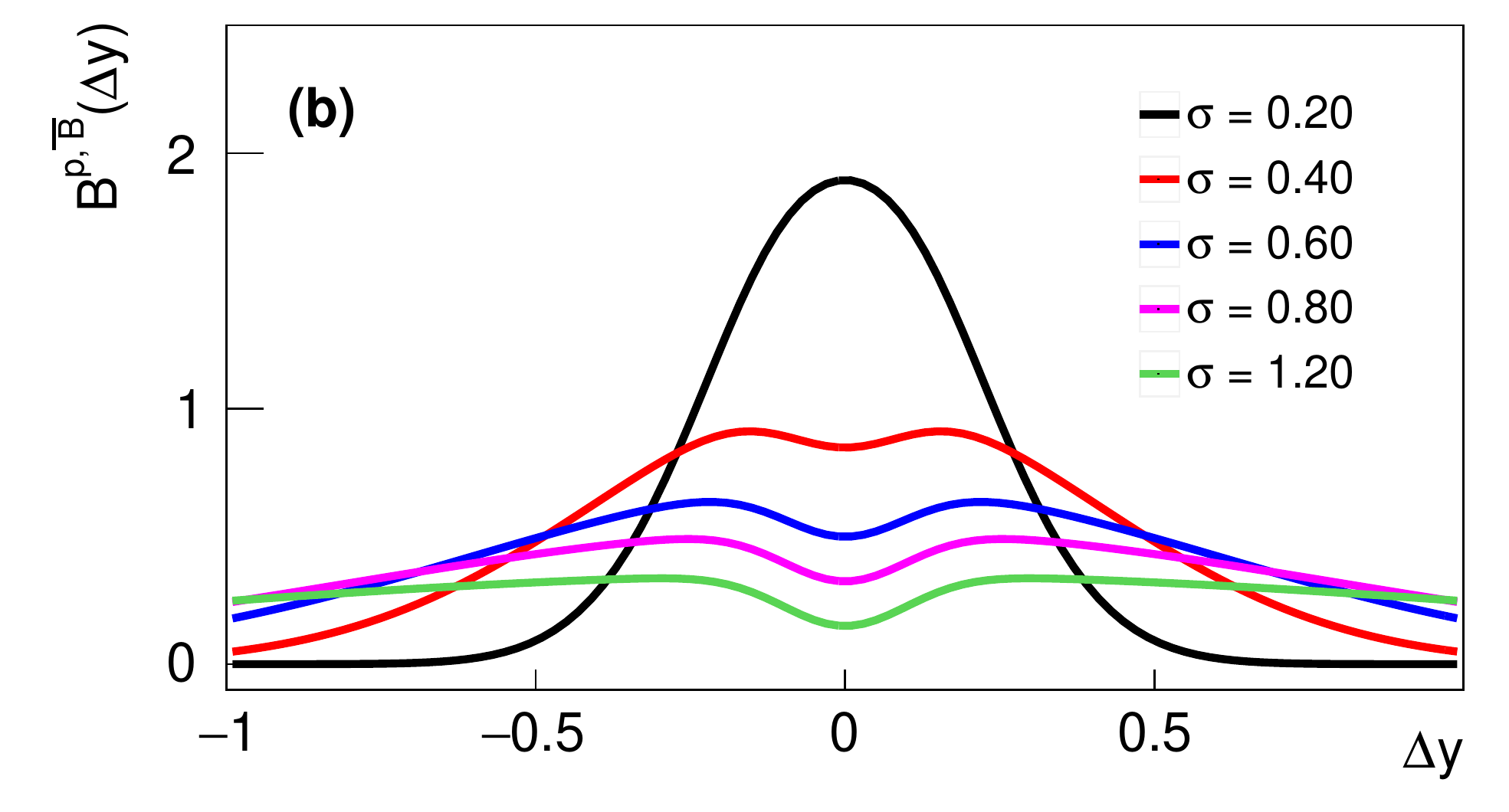} 
   \includegraphics[width=0.32\linewidth,keepaspectratio=true,clip=true,trim=10pt 0pt 5pt 5pt]{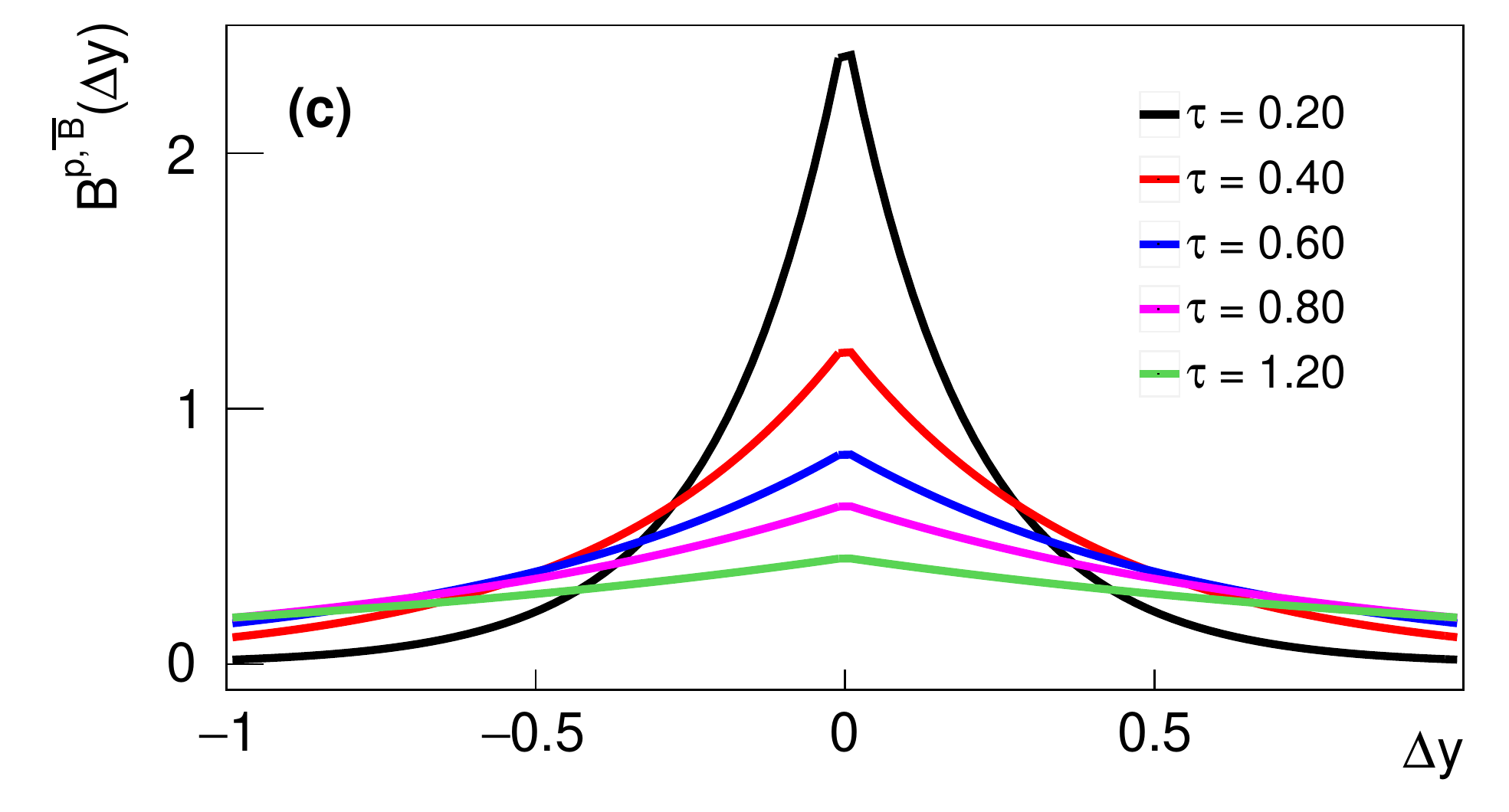} 
   \includegraphics[width=0.32\linewidth,keepaspectratio=true,clip=true,trim=10pt 0pt 5pt 5pt]{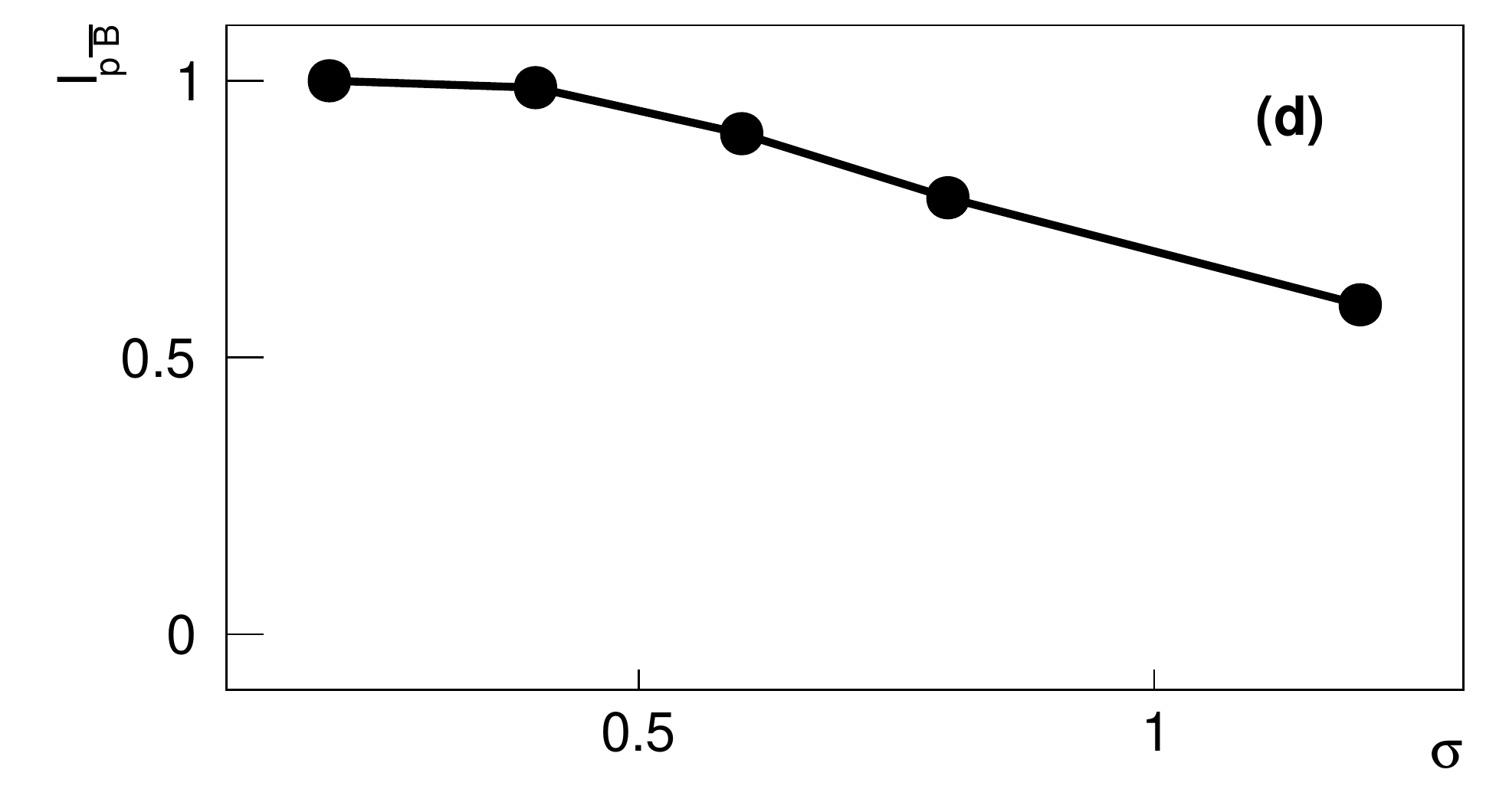} 
   \includegraphics[width=0.32\linewidth,keepaspectratio=true,clip=true,trim=10pt 0pt 5pt 5pt]{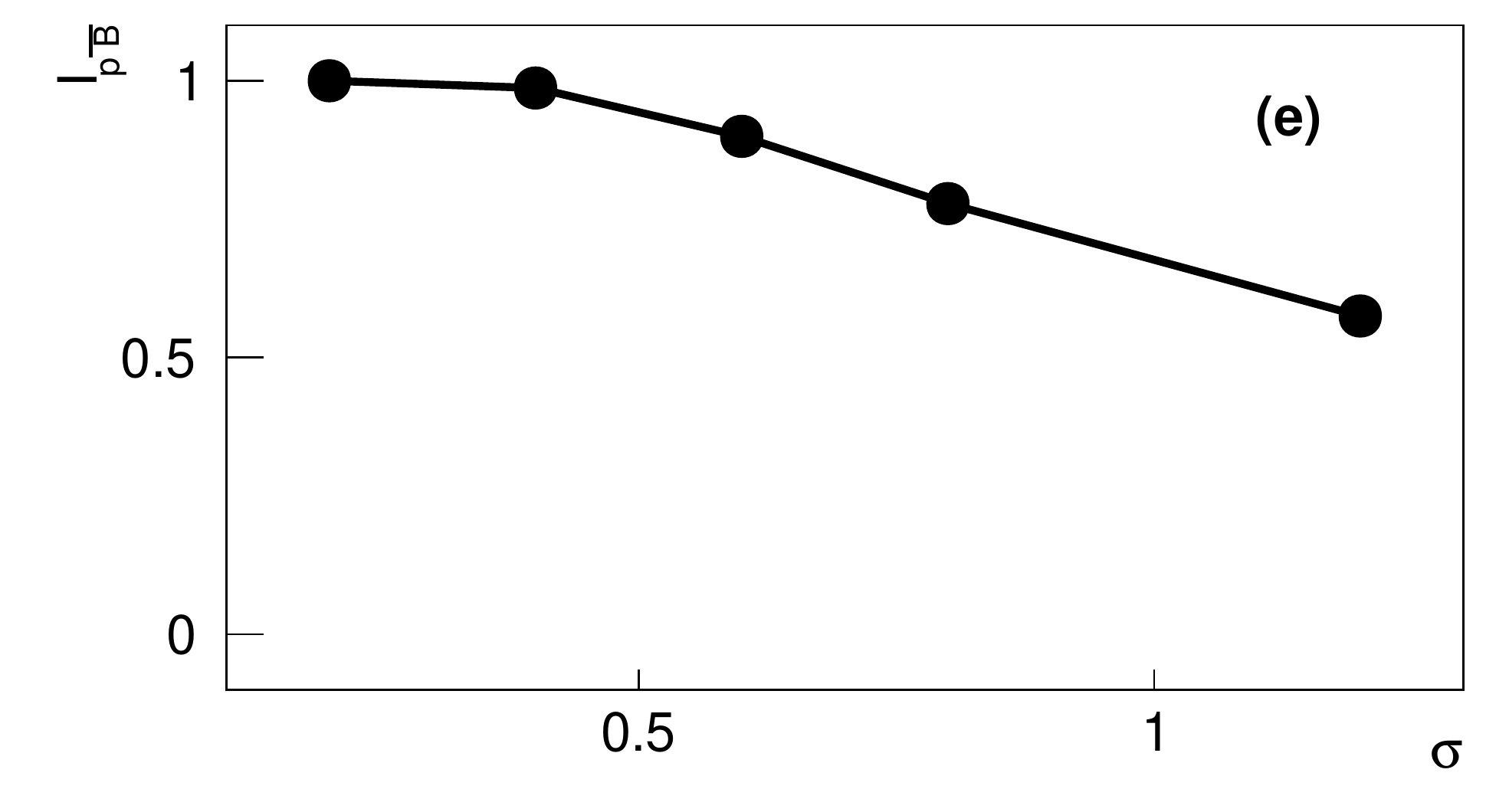} 
   \includegraphics[width=0.32\linewidth,keepaspectratio=true,clip=true,trim=10pt 0pt 5pt 5pt]{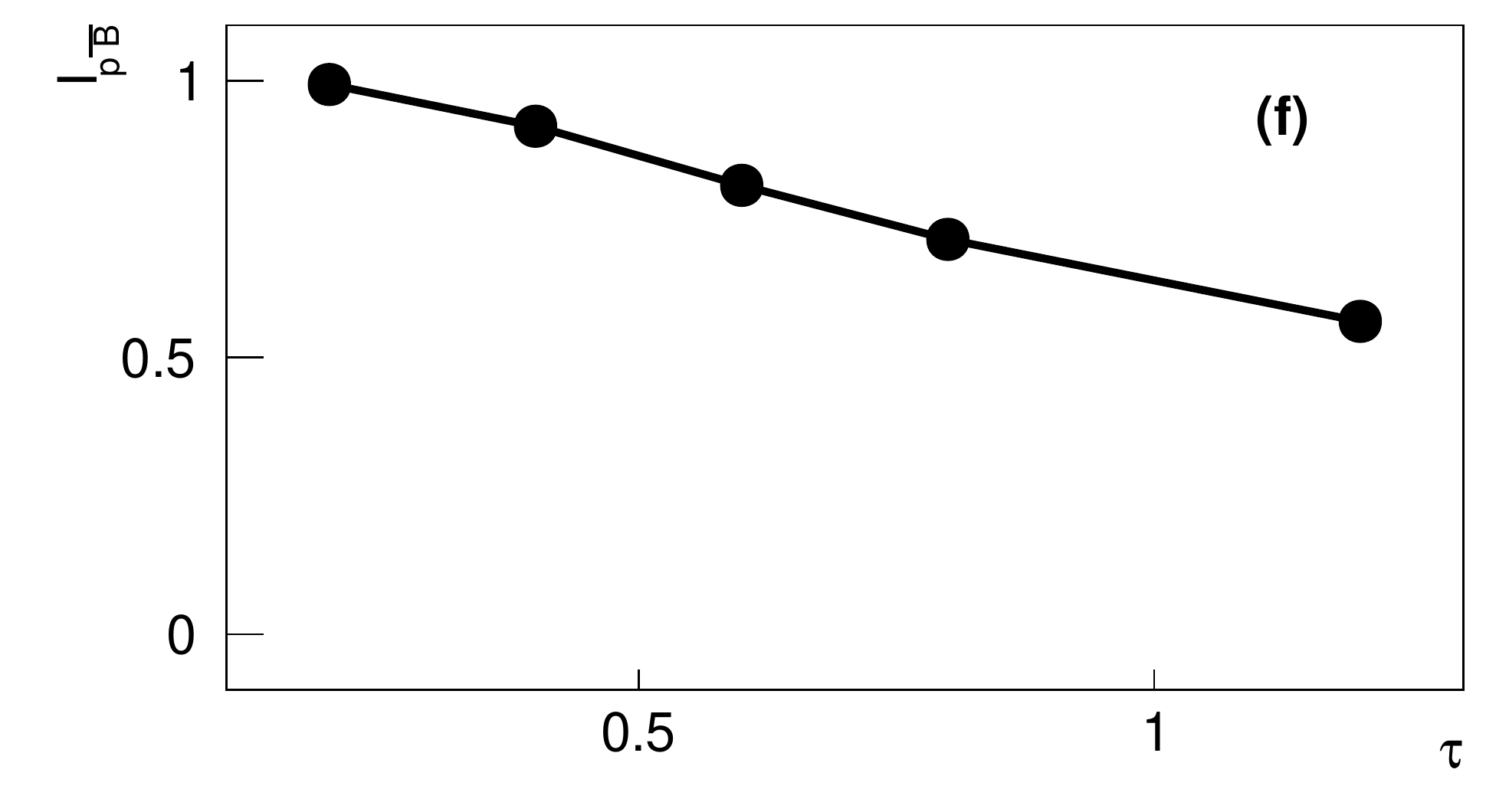} 
   \includegraphics[width=0.32\linewidth,keepaspectratio=true,clip=true,trim=10pt 0pt 5pt 5pt]{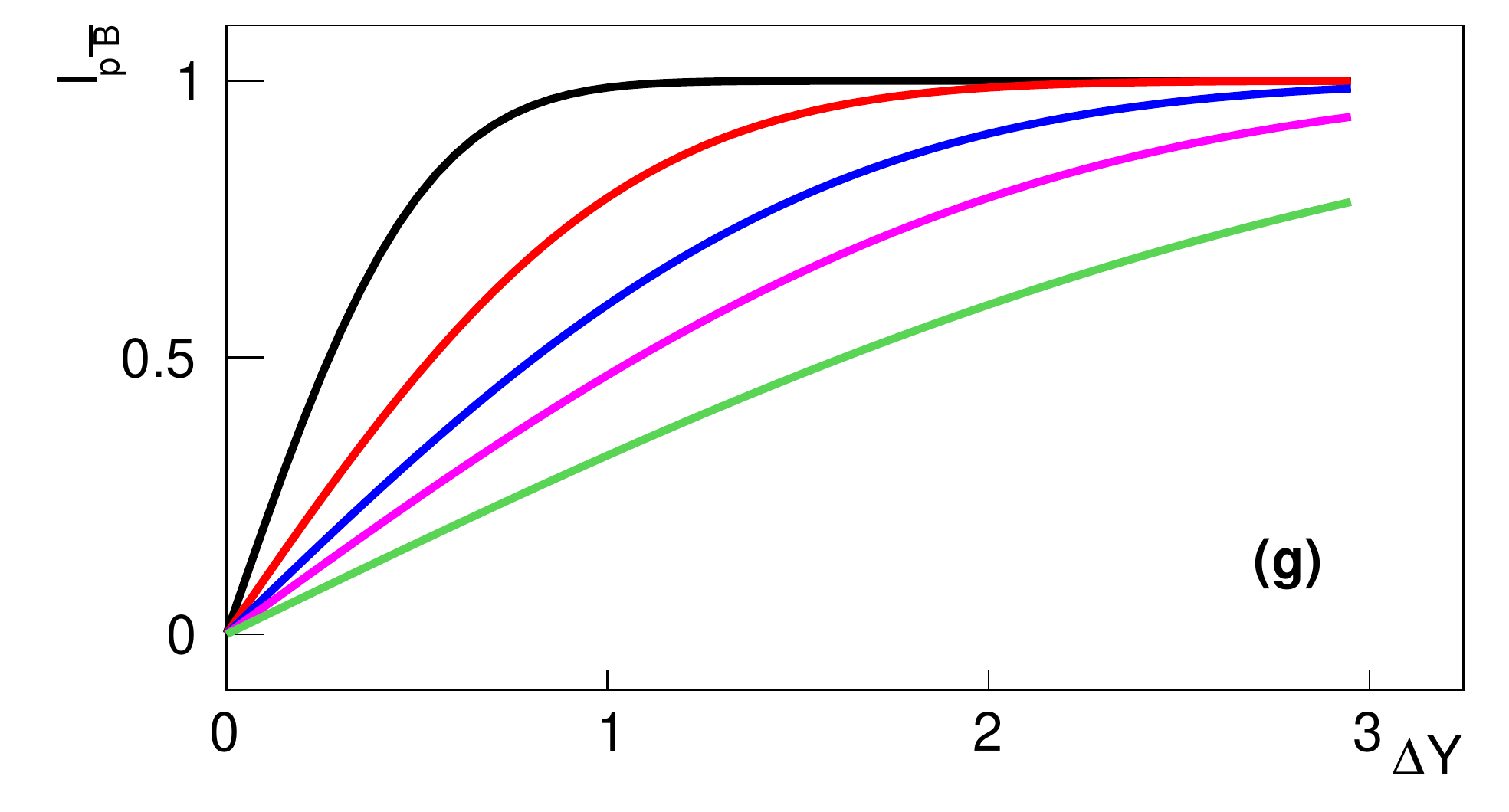} 
   \includegraphics[width=0.32\linewidth,keepaspectratio=true,clip=true,trim=10pt 0pt 5pt 5pt]{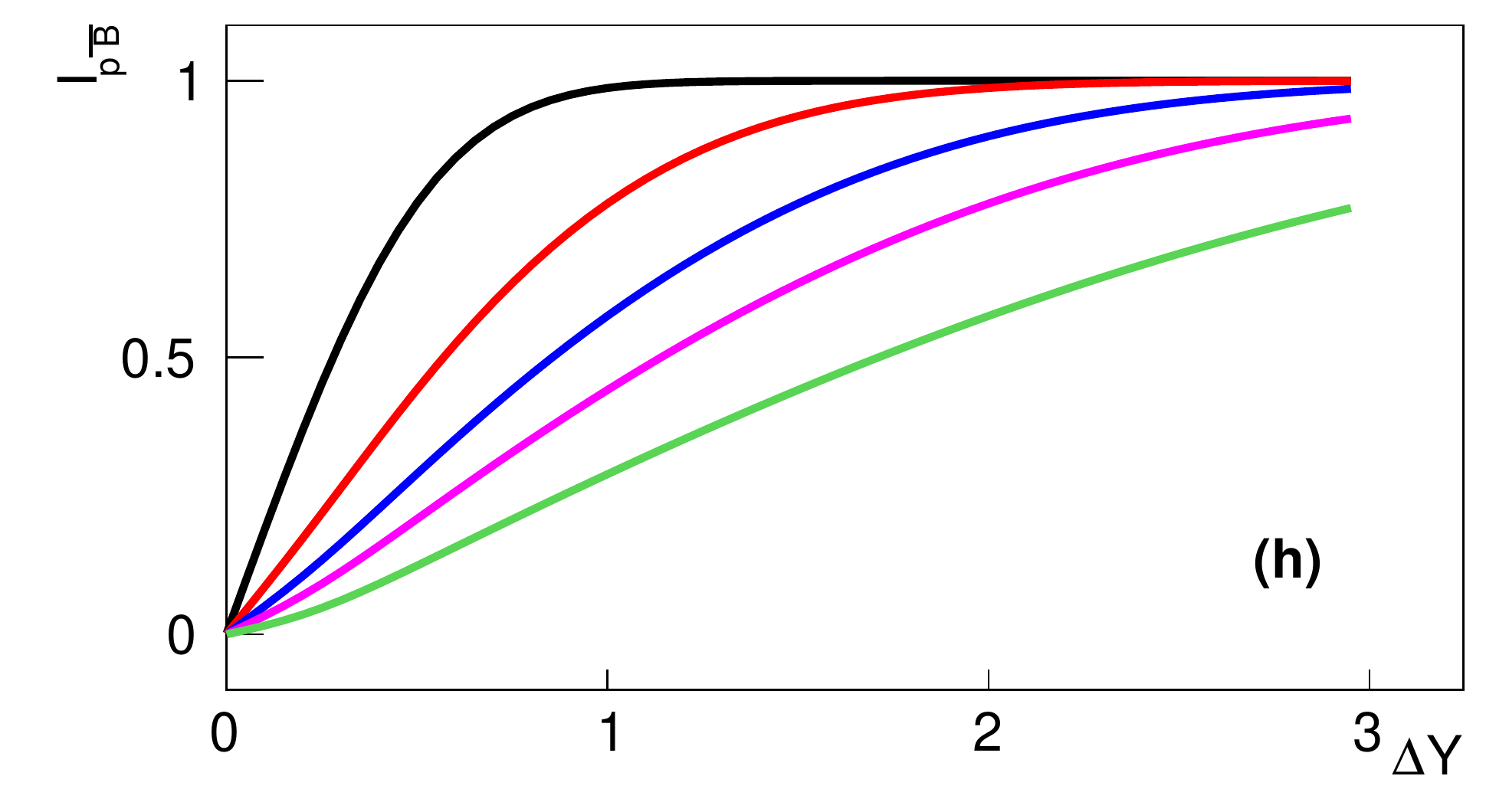} 
   \includegraphics[width=0.32\linewidth,keepaspectratio=true,clip=true,trim=10pt 0pt 5pt 5pt]{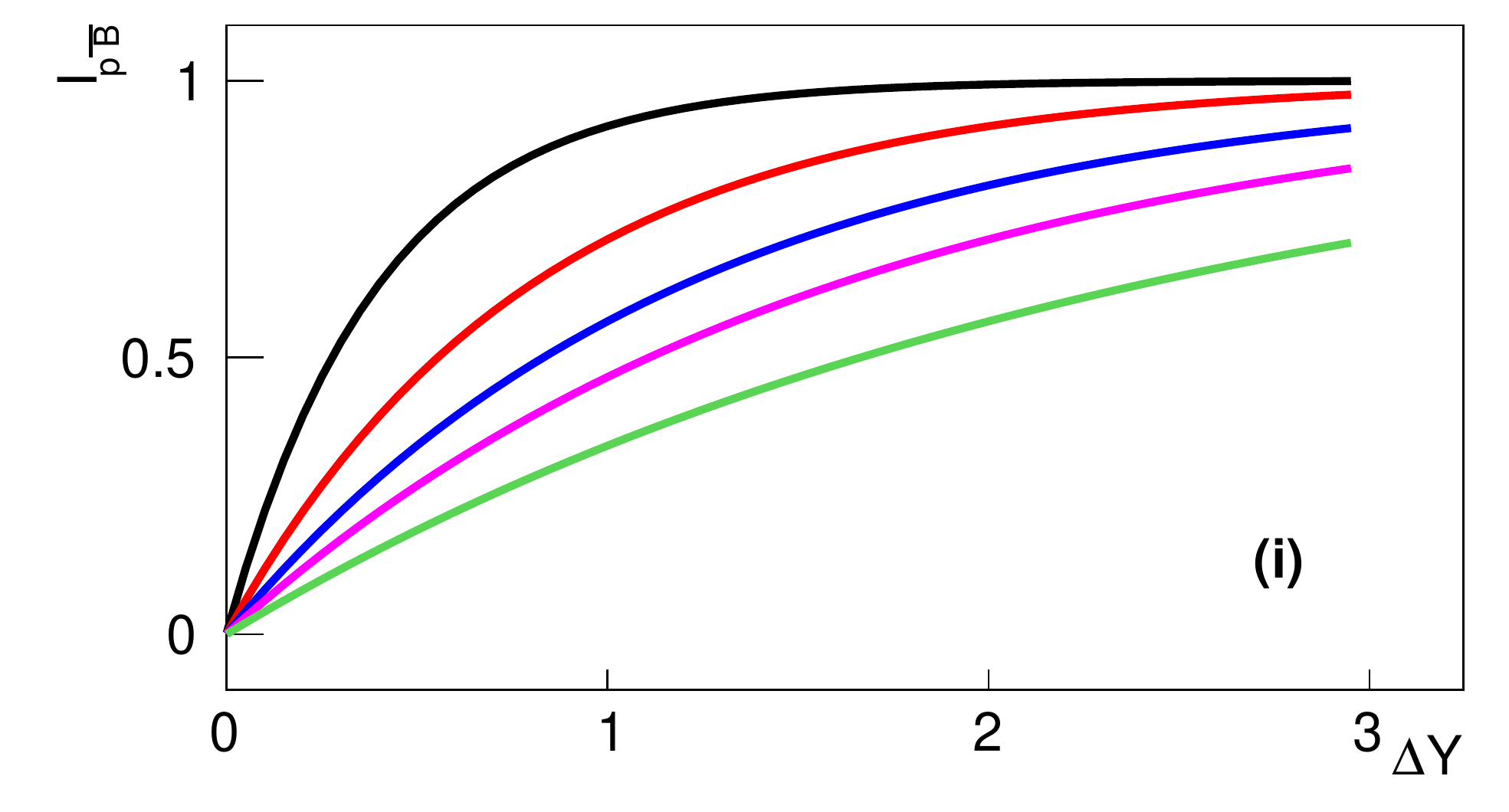} 
   \includegraphics[width=0.32\linewidth,keepaspectratio=true,clip=true,trim=10pt 0pt 5pt 5pt]{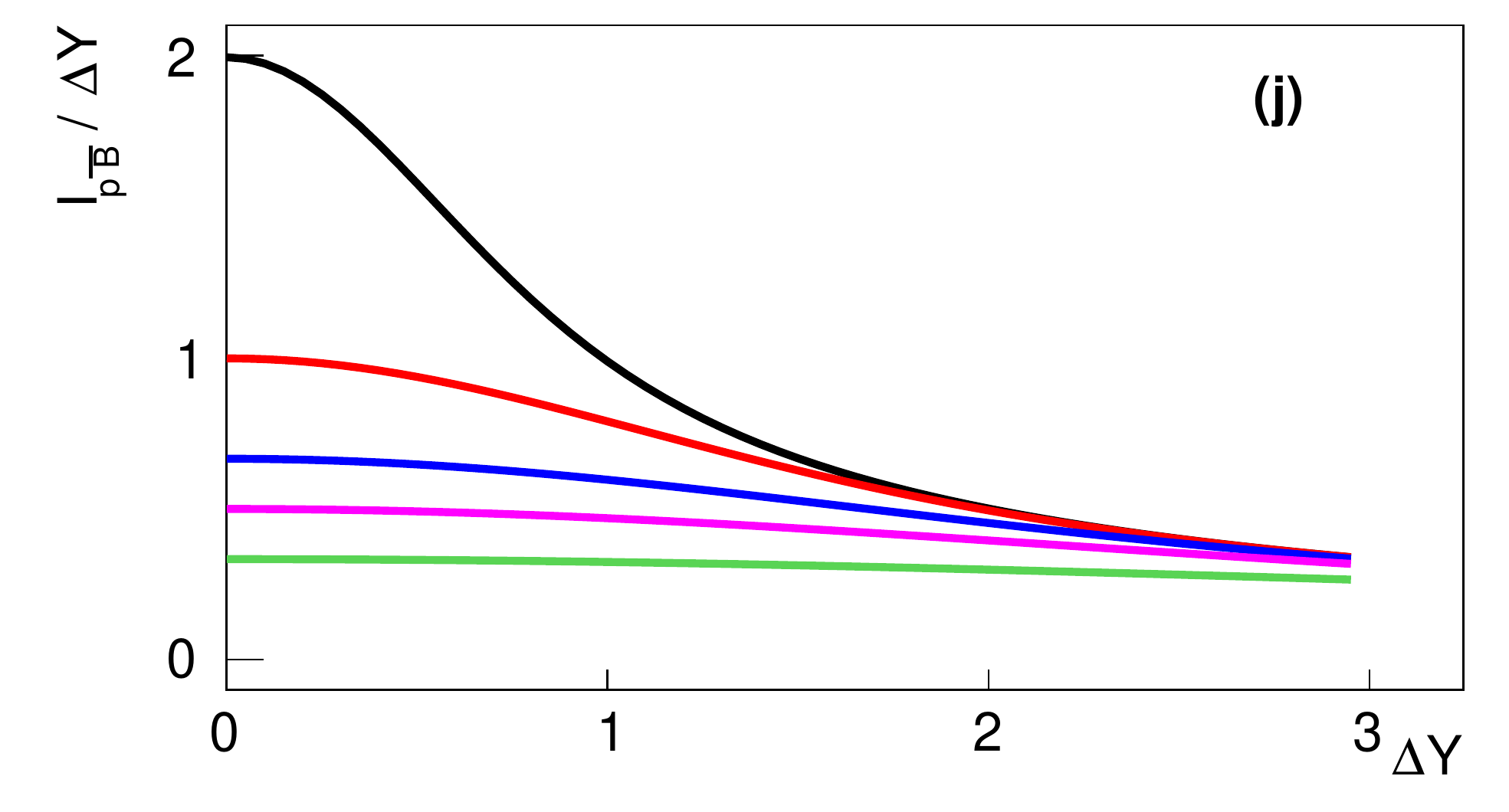} 
   \includegraphics[width=0.32\linewidth,keepaspectratio=true,clip=true,trim=10pt 0pt 5pt 5pt]{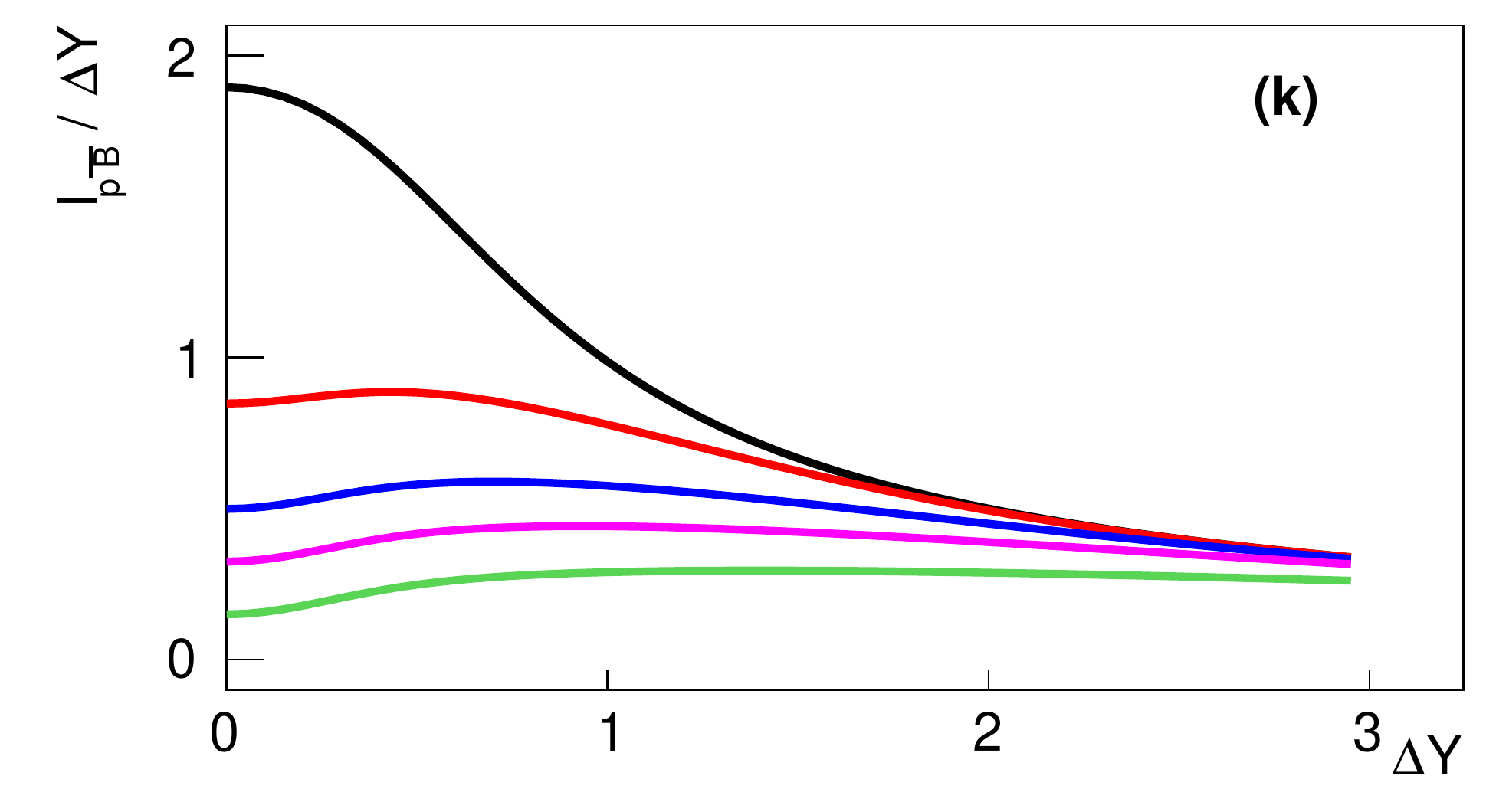} 
   \includegraphics[width=0.32\linewidth,keepaspectratio=true,clip=true,trim=10pt 0pt 5pt 5pt]{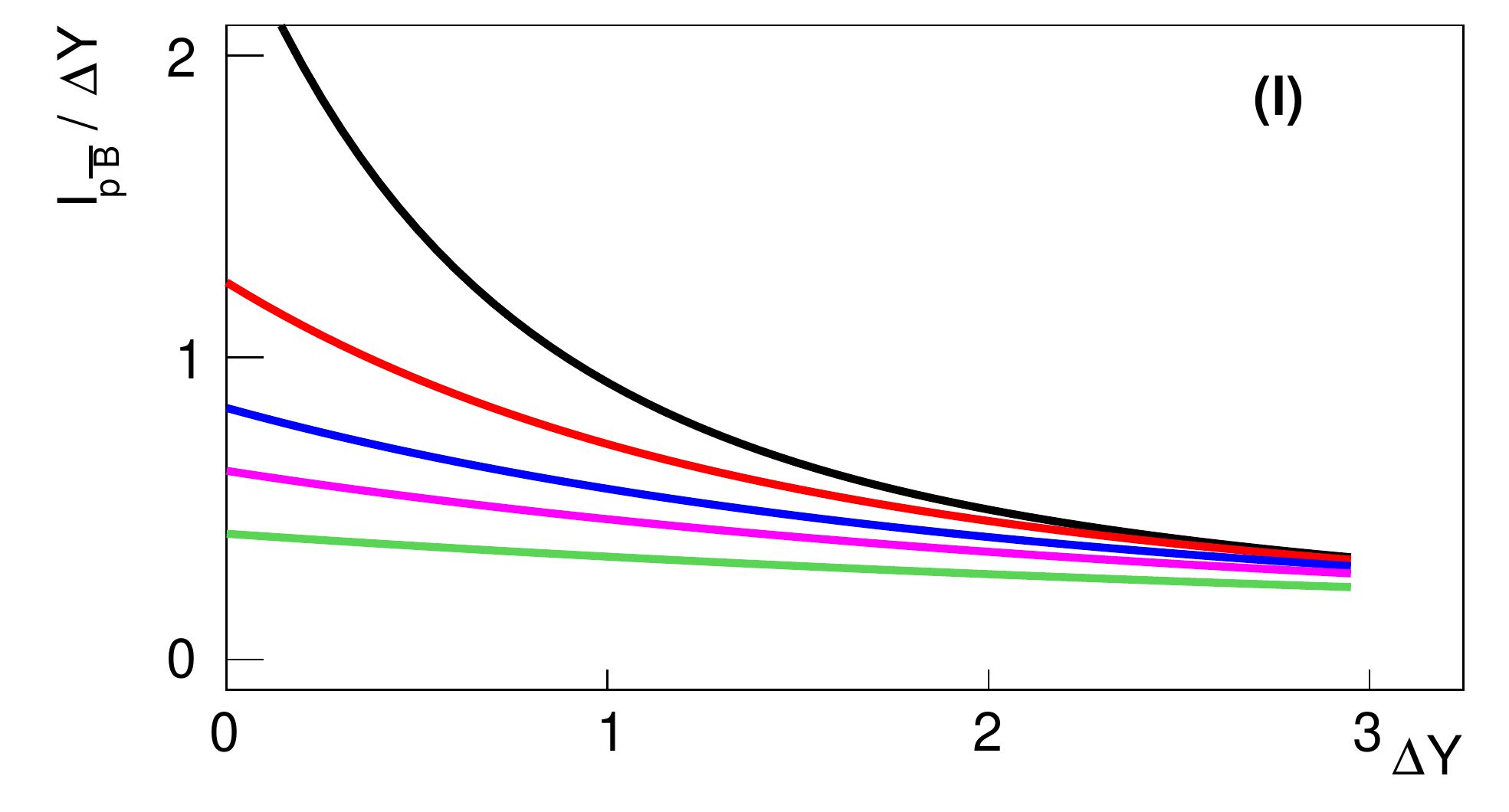} 
   \caption{Top row: (a) gaussian, (b) double-gaussian, and (c) exponential balance function models plotted as a function of the pair separation, $\Delta y=y_1-y_2$, for selected parameter values;   Upper middle row: Integrals of the (d) gaussian, (e) double gaussian, and (f) exponential   balance functions vs. the rms width ($\sigma$) or mean ($\tau$); Bottom middle row: Integrals $I_{p\bar B}(\Delta Y)$ of the (g) gaussian, (h) double gaussian, and (i) exponential   balance functions vs. the  width of the acceptance ($\Delta Y$) for models and parameter values shown in the top row. Bottom row: Ratio  $I_{p\bar B}(\Delta Y)/\Delta Y$ vs. $\Delta Y$.}
   \label{fig:BFexample}
\end{figure}
 
Experimentally, however, particles are measured within  limited (pseudo)rapidity and transverse momentum ranges. The probabilities  $I_{p,\bar \beta}^{4\pi}$,
are thus not directly measurable. Extrapolation of BF integrals to the full rapidity and momentum ranges of particle production are
non-trivial given they are highly dependent on their width and shape  (pair separation profile).  This is illustrated in 
Fig.~\ref{fig:BFexample}, which  presents examples of balance functions, with full $p_{\rm T}$ coverage,  and their respective integrals for  selected parameter values.  Panels (a-c) present  balance function with gaussian (G), double gaussian (DG), and exponential (E) dependence on the pair separation $\Delta y = y_1 - y_2$, respectively, and defined according to 
\be
B_{\rm G}(\Delta y) &=& \frac{1}{\sqrt{2\pi} \sigma} \exp\left(- \frac{\Delta y^2}{2\sigma^2} \right), \\ 
B_{\rm DG}(\Delta y) &=& \frac{1.05}{\sqrt{2\pi} \sigma} \exp\left( -\frac{ \Delta y^2 }{2\sigma^2} \right) - \frac{0.05}{\sqrt{2\pi} \sigma_{\rm N}} \exp\left( -\frac{\Delta y^2 }{2\sigma_{\rm N}^2} \right), \\ 
B_{\rm E}(\Delta y) &=& \frac{1}{ \tau} \exp\left( -\frac{|\Delta y|}{\tau} \right), 
\ee
where $\sigma$ is the rms width of the single gaussian rapidity, $\sigma_{\rm N}=0.1$ corresponds to the rms width of the narrow Gaussian used here to model, e.g., baryon annihilation, and $\tau$ is used to model the rate of decay of the rapidity density. 
Panels (d-f) presents integrals of the gaussian, double gaussian, and exponential balance function profiles as a function of the value of  $\sigma$ ($\tau$) for a  nominal acceptance $-1 < y <  1$.  The examples shown clearly  illustrate that  the integral $I_{p,\bar \beta}$ depends on the shape and width of the balance function  
relative to the measurement acceptance. This is further illustrated in panels (g-i), which display  integrals $I_{p\bar B}(\Delta Y)$ of the BFs shown in panels (a-c), as a function of  $\Delta Y = y_{\max}-y_{\min}$ denoting the width of the single particle acceptance $y_{\min} < y <  y_{\max}$.  One finds, indeed, that the rate at which the measured integral $I_{p,\bar \beta}$ converges to  its $4\pi$ limit is dependent on the shape of 
the balance function as well as its width $\sigma$. Also note that the  linear dependence on $\Delta Y$ expected from Eq.~(\ref{eq:rVsDeltaEta}) breaks down for these semi-realistic balance functions, as clearly illustrated by the plots of $I_{p,\bar B}/\Delta Y$ vs. $\Delta Y$ shown in the bottom  row of Fig.~1. For most cases considered, the ratio $I_{p,\bar B}/\Delta Y$  varies  with $\Delta Y$. The linear dependence  embodied in Eq.~(\ref{eq:rVsDeltaEta}) is thus indeed a poor approximation
of the actual dependence of $I_{p,\bar B}$ on the width $\Delta Y$ (or $\Delta \eta$) of the acceptance. 

\begin{figure}[htbp] 
   \centering
   \includegraphics[width=0.4\linewidth]{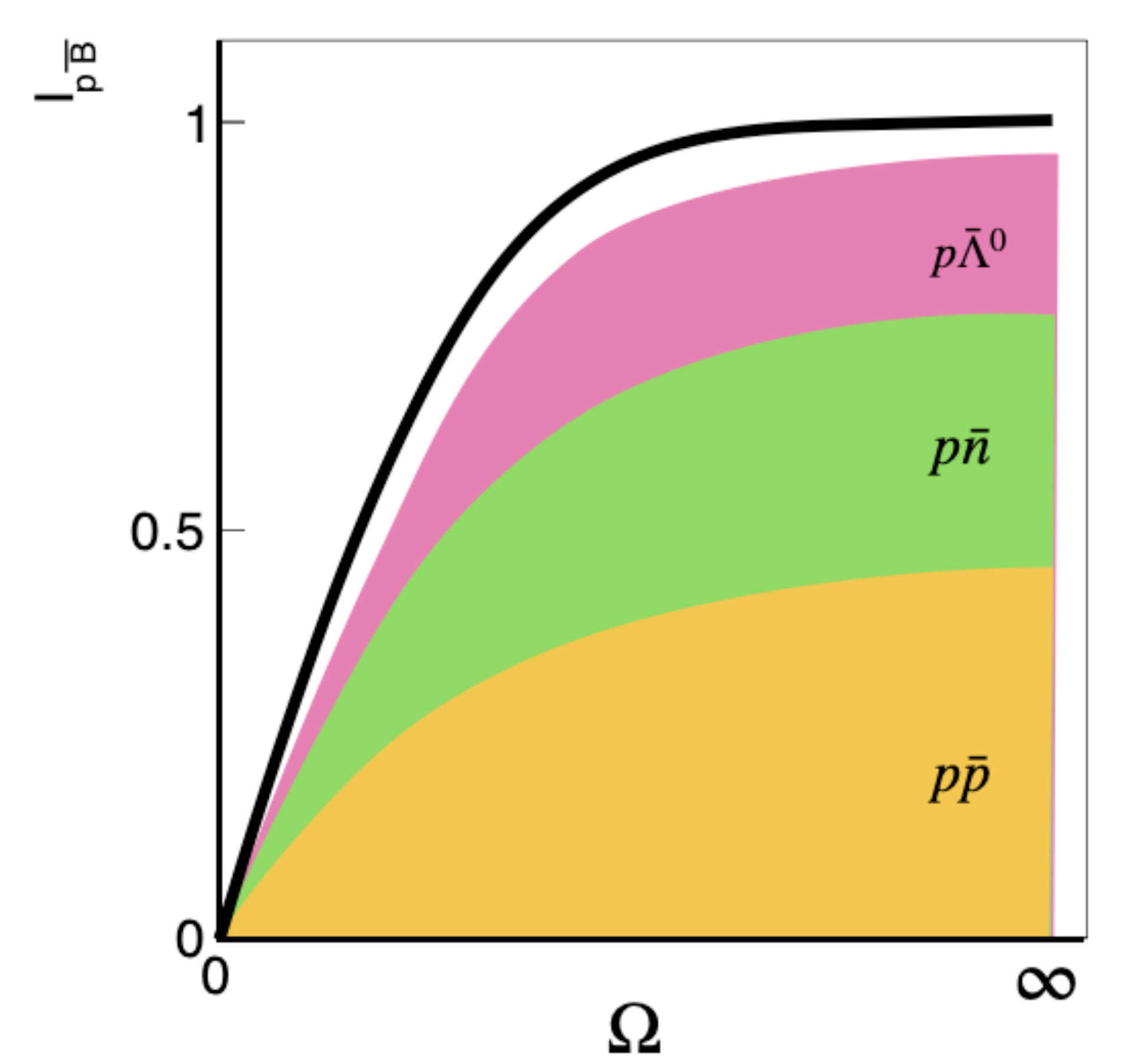} 
   \caption{Schematic dependence of the integral of  balance functions $B^{p,\bar B}(\Delta y)$ vs. the width of the experimental acceptance $\Omega$. The  colored bands schematically illustrate  contributions from distinct baryon number balancing anti-baryons.}
   \label{fig:example2}
\end{figure}

The above examples clearly  illustrate that the integral of the BF is a  function of its shape as well as the width $\Delta Y$ of the experimental acceptance. Given the baryon number balancing of the proton may be achieved  with several distinct anti-baryon species,  one must then consider the evolution of integrals $I_{p,\bar \beta}$ for all species $\bar \beta$ as a function of the 
measurement acceptance $\Omega$, as illustrated schematically in Fig.~\ref{fig:example2}. 

Next recall that the integral of the balance function is proportional to $\nu_{\rm dyn}^{p,\bar p}$ which, as we saw in Eq.~(\ref{eq:OneMinusR}), is also proportional to $1-r_{\Delta N_p}$. The magnitude of   $\kappa_2(\Delta p)$, measured at high energy, 
 is thus entirely determined by the integral of the balance function across the fiducial acceptance. The integral of the balance function, in turn, is determined by baryon number  conservation and the chemistry of the collision, i.e., what fraction of protons are accompanied by an anti-proton. If protons were balanced exclusively by anti-protons, the integral of the balance function over the entire phase space would yield unity. With finite ranges in $p_{\rm T}$ and rapidity $y$, the integral is determined 
by the width of these ranges. The larger they are, the closer the integral gets to saturation (unity if only anti-protons balance protons). The measured values of $\kappa_2(\Delta p)$ at LHC and top RHIC energy are thus determined ab-initio by baryon number conservation and the width of the balance function relative to that of the  acceptance. 

It is well established that the shape and width of the balance function of charge particles 
exhibit a significant narrowing with increasing collision centrality~\cite{PhysRevC.82.024905,Wang:2012jua,Abelev:2013csa}. This narrowing is understood to result largely from radial flow and was successfully modeled with the blast wave model: the more central collisions are,
the faster is the radial flow~\cite{Bozek2005247}. The value of $1-r_{\Delta N_p}$ is thus determined in large part by the magnitude 
of radial flow and the width of the acceptance and much less by the full coverage integral $I_{p,\bar \beta}^{4\pi}$.

Nominally, if effects of radial flow were invariant with collision centrality, the multiplicity $\la N_T\ra_{AA}$ measured in A--A collisions would scale in proportion to its value in pp collisions $\la N_T\ra_{pp}$ according
\be
\la N_T\ra_{AA} = \la n_s\ra \la N_T\ra_{pp},
\ee
where $\la n_s\ra$ is the effective number of sources involved, on average, in a given  A--A centrality range.
In contrast, one also expects that, in the absence of re-scattering of secondaries, that $\nu_{\rm dyn}^{p,\bar p (AA)}$
measured in A--A should scale as 
\be
\nu_{\rm dyn}^{p,\bar p (AA)} = \frac{1}{\la n_s\ra} \nu_{\rm dyn}^{p,\bar p (pp)},
\ee
relative to the value $\nu_{\rm dyn}^{p,\bar p (pp)}$ measured in pp collisions~\cite{Pruneau:2002yf}. Such
scaling is in fact essentially observed in Au--Au and Pb--Pb collisions~\cite{PhysRevC.68.044905,PhysRevC.79.024906,Abelev:2012pv}.  In this context, 
the ratio $r_{\Delta N_p}$ would then be invariant with A--A collision centrality. But the radial flow 
velocity is known to increase in more central collisions thereby leading to a narrowing of the balance function~\cite{PhysRevC.82.024905}. 
This consequently leads to  an increase of the  integral $I_{p,\bar \beta}$ within the experimental acceptance. The centrality 
dependence of $r_{\Delta N_p}$ shall then be driven primarily by the evolution of radial flow 
with collision centrality and it might have essentially nothing to do with the chemistry of the system and its susceptibility $\hat \chi_2^B$.

The width of the net-charge balance function is also observed to increase monotonically with decreasing beam energy ($\sqrt{s_{\rm NN}}$)~\cite{Adamczyk:2015yga}.  
This can be in part understood as a result of slower radial flow profile with decreasing beam energy. Should
the $p\bar p$ balance function behave in a similar fashion, one would expect the integral $I_{p,\bar p}$
to reduce monotonically with decreasing beam energy because the fraction of the BF within the acceptance
shrinks as its width increases. Once again, one expects the magnitude of $\kappa_2(\Delta p)$
to change with beam energy for reasons completely independent of the susceptibility $\hat \chi_2^B$.

However,  the ratio $\la N_{\bar p}\ra/\la N_p\ra$ is also known to fall rapidly with decreasing beam energy. The
$\la N_p\ra =  \la N_{\bar p}\ra$ hypothesis used to derive Eqs.~(\ref{eq:SkellamRatio},\ref{eq:OneMinusR}) is thus indeed strictly invalid 
at the low energy end of the BES. One must  thus examine the effect of baryon stopping 
on the fluctuations.

\section{Net Protons Fluctuations in the presence of nuclear stopping} 
\label{sec:Stopping}

In order to model the effect of baryon stopping, I will assume, as in \cite{PhysRevC.68.014907}, that  one can partition the measured 
protons into two subsets: the first, denoted $i$, corresponding to ``stopped" protons, and the second,
denoted $p$, corresponding to protons produced by $p\bar B$ pair creation. All anti-protons are 
assumed produced by pair production and I will neglect, for simplicity, the impact of annihilation. 

I thus consider  Eq.~(\ref{eq:kapp2Skellam2}) with the following substitutions
for the first and second order factorial cumulants of protons and anti-protons:
\be \nonumber
F_1^{p} &\rightarrow& F_1^{i} + F_1^{p} = \la N_i\ra + \la N_p\ra \\  \nonumber
F_1^{\bar p} &\rightarrow& F_1^{\bar p} =  \la N_{\bar p}\ra\\ 
F_2^{p,p} &\rightarrow& F_2^{i,i} + F_2^{i,p} + F_2^{p,i} + F_2^{p,p} \\ \nonumber
F_2^{p,\bar p} &\rightarrow& F_2^{i,\bar p} +  F_2^{p,\bar p} \\ \nonumber
F_2^{\bar p,p} &\rightarrow& F_2^{\bar p,i} +  F_2^{\bar p,p} \\ \nonumber
F_2^{\bar p,\bar p} &\rightarrow& F_2^{\bar p,\bar p}.  \nonumber
\ee  
In symmetric A--A collisions, one must have $F_2^{i,p}=F_2^{p,i}$,  $F_2^{i,\bar p}=F_2^{\bar p,i}$, and $F_2^{p,\bar p} = F_2^{\bar p,p}$.   
Introducing $\la N_T\ra = \la N_i\ra +  2\la N_p\ra$ and $\xi = \la N_i\ra/\la N_T\ra= F_1^{i}/\left(F_1^{i}+2F_1^{p}\right)$, one gets    
\be\label{eq:BES}
r_{\Delta N_p} &=& 1 \hspace{-0.04in}+  \hspace{-0.04in}
\Scale[1]{\frac{ F_2^{i,i} + 2 F_2^{i,p} + F_2^{p,p} + F_2^{\bar p,\bar p} -2F_2^{i,\bar p}  - 2F_2^{p,\bar p} }
{F_1^{i} + F_1^{p}+ F_1^{\bar p}}}, \\ 
\label{eq:BES2}
&=& 1 
+\xi^2 \la N_T\ra R_2^{i,i}   + \frac{1}{4} \left(1-\xi \right)^2 \la N_T\ra  \nu_{\rm dyn}^{p,\bar p},
\ee
where in the second line, I neglected effects of annihilation, which imply that  $F_1^{p} = F_1^{\bar p}$,  $F_2^{p,p}=F_2^{\bar p,\bar p}$, and I assumed $F_2^{i,p} \approx F_2^{i,\bar p}$.
The second term, proportional to $R_2^{i,i}$,  is a measure of the correlation strength  of stopped protons, while the third term, proportional to  $\nu_{\rm dyn}^{p,\bar p}$ corresponds to the pair creation component found in the high-energy limit, Eq.~(\ref{eq:SkellamRatio}).  Experimentally, it has been observed that 
nucleons from the projectile and target lose, on average, approximately two units of rapidity  in nuclear collisions. 
At LHC and top RHIC energy, this leads to a vanishing net-baryon density in the central rapidity region
but for decreasing  $\sqrt{s_{\rm NN}}$, and particularly at the low end of RHIC the beam energy scan, 
this yields a large net  proton excess at central rapidity. Given the production of $p\bar p$ pairs is a logarithmic function 
of $\sqrt{s_{\rm NN}}$, one expects the term proportional to $R_2^{i,i}$ should largely dominate
at the low end of the BES range while the term proportional to $\nu_{\rm dyn}^{p,\bar p}$, driven
by baryon number conservation, should dominate at LHC and top RHIC energy.  Equation (\ref{eq:BES2}) 
thus tells us that  the beam energy evolution 
of $r_{\Delta N_p}-1$ should be determined by the interplay of baryon stopping and net-baryon conservation, the former and the latter dominating at  low and high  $\sqrt{s_{\rm NN}}$, respectively. Given the strength and $\Delta y$ dependence of  $R_2^{i,i}(\Delta y)$ and 
$\nu_{\rm dyn}^{p,\bar p}(\Delta y)$ are determined by different mechanisms, they  shall likely have distinct dependences on $\sqrt{s_{\rm NN}}$. As the contribution of stopped baryons decreases with 
increasing $\sqrt{s_{\rm NN}}$, one  thus anticipates that the  balance function of created pairs $p,\bar p$, and thus $\nu_{\rm dyn}^{p,\bar p}(\Delta y)$),  shall 
dominate. The net-proton fluctuations   $r_{\Delta N_p}-1$ might then exhibit a rather complicated dependence on $\sqrt{s_{\rm NN}}$. Such a dependence, however, has little to do with the properties of nuclear matter near equilibrium and more to do 
with dynamic considerations including nuclear stopping power and radial flow resulting from large 
inside-out pressure gradients.

\section{Net Baryon Fluctuations} 
\label{sec:NetBaryonFluctuations}

Equation (\ref{eq:susceptibility}) relates the baryon susceptibility $\hat \chi^B_2$ to the second cumulant of the net baryon number $\Delta B$. One must thus consider, at least in principle, the fluctuations of all baryons and anti-baryons, $\Delta B = N_B - N_{\bar B}$, not only those of  the net proton number 
$\Delta N_p$.  Repeating the derivation presented in Sec.~\ref{sec:SkellamLimit} for net-baryon fluctuations, one gets 
\be\label{eq:kapp2SkellamBaryon}
r_{\Delta B} \equiv \frac{\kappa_2(\Delta N_B)}{\kappa_2^{\rm Skellam}(\Delta N_B)}  = 1+\frac{F_2^{B,B} +  F_2^{\bar B,\bar B} - 2 F_2^{B,\bar B} }{F_1^B+F_1^{\bar B}},
\ee
which, in the high-energy limit, yields 
\be
r_{\Delta B}-1 &=& \frac{1}{4}\la N_{TB} \ra  \nu_{\rm dyn}^{B,\bar B} =  I_{B,\bar B}(\Omega),
\ee
where $\la N_{TB} \ra = \la N_{B} \ra + \la N_{\bar B} \ra$ and $I_{B,\bar B}(\Omega)$ is the integral of the baryon--baryon balance function $B^{B,\bar B}$. 

In order 
to express $B^{B,\bar B}$ in terms of elementary balance functions $B^{\alpha,\bar \beta}$, first note that single- and two-baryon densities 
can written
\be
\rho_1^{B} = \sum_{\alpha}   \rho_1^{\alpha};  \hspace{0.3in}
\rho_2^{B,B} = \sum_{\alpha}  \sum_{\beta} \rho_2^{\alpha,\beta},
\ee
where sums on $\alpha$ and $\beta$  span all produced baryons. Similar expressions can be written for single- and pair-densities involving anti-baryons.
Defining the yield fractions 
\be
f_{\alpha} = \frac{\rho_1^{\alpha}}{\rho_1^{B}};  \hspace{0.3in}f_{\bar\alpha} = \frac{\rho_1^{\bar\alpha}}{\rho_1^{\bar B}},
\ee
such that $\sum_{\alpha}f_{\alpha} =1$ and $\sum_{\bar \alpha}f_{\bar\alpha} =1$, one finds that the baryon-baryon balance function $B^{B,\bar B}$ may be written
\be
B^{B,\bar B}(\Delta y)= \frac{1}{2}\left[ \sum_{\alpha} f_{\alpha}  D_2^{\alpha,\bar B }(\Delta y) +  \sum_{\bar\alpha} f_{\bar\alpha}  D_2^{\bar\alpha,B}(\Delta y) \right].
\ee
In the high-energy limit, one has $f_{\alpha}= f_{\bar\alpha}$, and the above expression simplifies to 
\be
B^{B,\bar B}(\Delta y)= \sum_{\alpha} f_{\alpha} B^{\alpha,\bar B}(\Delta y),
\ee
where 
\be
B^{\alpha,\bar B}(\Delta y) = \sum_{\bar\beta} B^{\alpha,\bar \beta}(\Delta y).
\ee
Single particle production yields measured in heavy-ion collisions are very well described in the context of thermal production models 
determined by a (chemical) freeze-out temperature as well as  charge and strangeness chemical potentials.  Within such models, 
one finds the baryon (anti-baryon) production is dominated by the lowest mass states (e.g., proton, neutron). The baryon-baryon balance function, $B^{B,\bar B}(\Delta y)$, shall 
thus be dominated by contributions from proton-baryon, $B^{p,\bar B}(\Delta y)$, neutron-baryon, $B^{n,\bar B}(\Delta y)$,  balance functions, with  weaker
contributions from $\Lambda$-baryon or heavier strange baryons and with negligible contributions from charm or bottom baryons.
On general grounds, and neglecting electric charge (or isospin), one can expect $B^{p,\bar B}(\Delta y)$ and  $B^{n,\bar B}(\Delta y)$ to feature similar strengths and dependence on $\Delta y$.    However, balance functions involving strange baryons, in particular  $B^{p,\bar \Lambda}(\Delta y)$ and $B^{p,\bar \Sigma}(\Delta y)$, might have a rather different dependence on $\Delta y$ owing to the fact 
that s-quarks may be produced at earlier times than u- and d-quarks, or be subjected to different transport mechanisms. Fortunately, 
measurements of $B^{p,\bar \Lambda}(\Delta y)$, $B^{\Lambda,\bar \Lambda}(\Delta y)$, and perhaps even $B^{p,\bar \Sigma}(\Delta y)$, are in principle possible. One can then anticipate, in a near future, being able to estimate the shape and strength of $B^{p,\bar B}(\Delta y)$ and $B^{B,\bar B}(\Delta y)$ based on measurements within the acceptance 
of ongoing experiments (e.g., ALICE). 

\section{Summary} 
\label{sec:Summary}

I showed there is straightforward connection between the fluctuations of net-baryon number measured
at central rapidities in A--A collisions in terms of second order cumulants of the net-baryon number
and the strength of two-particle correlations factorial cumulants. I further showed that in the high-energy
limit, corresponding to a vanishing net-baryon number, fluctuations are entirely determined by the strength and
width of the $p\bar p$ balance function relative to the width of the acceptance.   By contrast, at low energy, the fluctuations of the net-baryon number are more likely
dominated by proton-proton correlations resulting from nuclear stopping. Overall, one 
can expect the fluctuations to display  a smooth evolution with $\sqrt{s_{\rm NN}}$ between
these two extremes but nowhere can one expect the magnitude of the fluctuations to be trivially 
sensitive to the nuclear matter baryon susceptibility $\chi_2^B$.

I here focused the discussion on 
second order cumulants of the net-baryon number but it is clear that the same line of argument can 
be extended to higher cumulants.  Measurements of fluctuations by STAR at RHIC have used
the magnitude of the second order cumulant of the net-baryon number as a reference
to factor out the ill defined notion of volume involved in relations between  cumulants
and susceptibilities. This would make sense if the susceptibilities  determined the magnitude of the
cumulants. But, as I have shown, the magnitude of $\kappa_2(\Delta p)$ is in fact
determined largely  by the width of the acceptance of the measurement relative to the width of the balance function
at high-energy and by proton-proton correlations associated with nuclear stopping at low energy. The use
of $\kappa_2(\Delta p)$ thus does not provide a sound basis to cancel out volume effects and
normalize the magnitude of higher cumulants. 

All is not lost, however. Measurements of momentum dependent balance functions may be used to quantitatively assess the role 
of both  baryon number conservation and nuclear stopping, and henceforth obtain sensitivity to QCD matter susceptibilties.
Additionally, measurements of  balance functions of pairs  $(p,\bar p)$, $(p,\bar \Lambda)$, $(\Lambda,\bar \Lambda)$,
and perhaps even $(p,\bar \Sigma)$, are possible. Results from these measurements shall inform  
our understanding of the system expansion dynamics,  our knowledge of the hadronization chemistry, and enable, as per the discussion 
in Sec.~\ref{sec:NetBaryonFluctuations},  an assessment of the relative strength of their contributions to net-baryon fluctuations. 
\newenvironment{acknowledgement}{\relax}{\relax}
\begin{acknowledgement}
\section*{Acknowledgements}

The author thanks colleagues S. Basu, C. Shen,  J. Pan, K. Read, and S. Voloshin   for fruitful discussions and their insightful comments. This work was supported in part by the United States Department  of Energy, Office of Nuclear Physics (DOE NP), United States of America under Award Number DE-FG02-92ER-40713.

\end{acknowledgement}

\bibliography{paper.bib}

\end{document}